\documentstyle[seceq,twoside]{ptptex}
\notypesetlogo  
\title{Light $\sigma$-Meson Production in Excited $\Upsilon$ 
Decay Processes II\\
--- Theoretical Investigation ---}
\author{%
Muneyuki {\sc Ishida}, Shin {\sc Ishida}$^*$ and Toshihiko {\sc Komada}$^*$
}
\inst{%
Department of Physics, Tokyo Institute of Technology\\
Tokyo 152-8551, Japan\\
$^*$ Atomic Energy Research Institute, 
College of Science and Technology\\
Nihon University, Tokyo 101-0062, Japan
}
\abst{%
The results of analyses obtained in Part I are examined from the viewpoint of chiral symmetry.
The mass spectra of $\pi\pi$ system are widely believed to be 
suppressed generally near the threshold, 
because of the derivative coupling property of Nambu-Goldstone 
$\pi$ meson. 
However, this suppression does not hold
in the processes with large energy release, 
and the steep increase from the $\pi\pi$ threshold observed in the transition
$\Upsilon (3S\to 1S)$ is shown to be consistent with general
constraints from chiral symmetry. 
}

\begin{document}
\maketitle

\setcounter{tocdepth}{4}

\section{Introduction}
In Part I,\cite{rf1} we have analyzed systematically the $\pi\pi$ production 
amplitudes in the transitions, $\Upsilon (2S)\to \Upsilon (1S)$,
$\Upsilon (3S)\to \Upsilon (1S)$, $\Upsilon (3S)\to \Upsilon (2S)$ and 
$\psi (2S)\to J/ \psi (1S)$, and the $\pi\pi$ and $KK$ production amplitudes in the trasition,
$J/\psi \to \phi$.
The production amplitudes ${\cal F}$ are parametrized 
in the form of the coherent sum of 
$\sigma$ Breit-Wigner amplitude ${\cal F}_\sigma$ and direct
$2\pi$ amplitude ${\cal F}_{2\pi}$, following the VMW method.
\begin{eqnarray}
{\cal F} &=& {\cal F}_\sigma + {\cal F}_{2\pi};\ \ 
{\cal F}_\sigma =\frac{r_\sigma e^{i\theta_\sigma}}{m_\sigma^2-s-i\sqrt s \Gamma_\sigma (s)},
\ \  {\cal F}_{2\pi}=r_{2\pi}e^{i\theta_{2\pi}}.
\label{eq1}
\end{eqnarray}
Here $r_\sigma$ ($r_{2\pi}$) is\footnote{
In our analysis the $r_\sigma$ and $r_{2\pi}$ are taken as free parameters,
since the $|\sigma\rangle$ state and $|2\pi\rangle$ state are considered,
from quark physical picture,\cite{rf7} as independent
bases of $S$-matrix, and have independent production couplings, in principle.  
} 
production coupling constant of $\sigma$-state 
($2\pi$-state) and $e^{i\theta_\sigma}$ ($e^{i\theta_{2\pi}}$) 
is a strong phase factor.
The $m_{\pi\pi}$ or $m_{KK}$ spectra are described well 
through all the relevant processes. 
Especially the double peak structure in $\Upsilon (3S)\to \Upsilon (1S)\pi\pi$ decay spectra
was nicely reproduced by the interference between ${\cal F}_\sigma$ and ${\cal F}_{2\pi}$.
The obtained $S$-matrix pole position of $\sigma$ state
is $m_\sigma -i\Gamma_\sigma /2=526-i150$MeV, which is taken 
commonly through all the relevant processes.
This seems to give a strong evidence for existence of $\sigma (500$--600).

However, we must give special attention on the threshold behaviors. 
Because of the derivative coupling property of $\pi$ meson as Nambu-Goldstone boson 
appearing in chiral symmetry breaking,  
the spectra of $|{\cal F}|^2$ is widely believed to be suppressed 
in the $\pi\pi$ low energy region.
For example, in the $\pi\pi$ scattering, the observed spectra is suppressed
near the threshold. 
In the linear $\sigma$ model this suppression is produced by a strong cancellation
between the $\sigma$ amplitude and the $\lambda\phi^4$ 
amplitude. In the relevant problem
this suppression is actually observed experimentally 
in $\Upsilon (2S\to 1S)$ and $\psi (2S\to 1S)$, and is reproduced, 
similarly 
by the cancellation between ${\cal F}_\sigma$ and ${\cal F}_{2\pi}$.
However, in $\Upsilon (3S)\to\Upsilon (1S)\pi\pi$ decay,
the steep increase from the $\pi\pi$ threshold is observed 
and is
reproduced by constructive interference between
${\cal F}_\sigma$ and ${\cal F}_{2\pi}$, seemingly to be inconsistent 
with the derivative coupling property of pion. 

The relevant decay processes had conventionally been treated 
as the intermediate two gluon emission process  
from transition between heavy quarkonium systems
(described by the multipole expansion\cite{rf2,rf3} of QCD), being accompanied by
the conversion of gluons into pions (described by current algebra and PCAC).
The main term of the amplitude, being proportional\cite{rf3} to $s$, comes from the trace of
the gluonic part of QCD energy momentum tensor $\theta_{\mu\mu}^G$, which
is enhanced due to the trace anomaly in QCD. The resulting amplitude has 
Adler 0
around $s\sim 0$, and predicts the suppression of the spectra 
near $\pi\pi$ threshold in the general processes,
while this is not valid in the transition 
$\Upsilon (3S)\to \Upsilon (1S)\pi\pi$, as was stated above.
This seems to show that the multipole expansion method is\footnote{
Multipole expansion method is effective in case $\langle kr \rangle\ll 1$, 
where $k$ is a typical momentum of the emitted gluon, which may be determined as
$k\approx (M'-M)/2$ ($M'(M)$ being mass of the initial (final) quarkonium).
$r$ is the size of the quarkonium, which is estimated, 
by using the quark model,\cite{rf5} with the values,\cite{rf6} 
$\langle r\rangle =3.5$GeV$^{-1}$ for $\Upsilon (3S)$
and  $\langle r\rangle =2.3$GeV$^{-1}$ for $\Upsilon (2S)$.
Thus, $\langle kr\rangle \approx 0.65$ for $\Upsilon (2S\to 1S)$, 
while  $\langle kr\rangle \approx 1.6$ for $\Upsilon (3S\to 1S)$.
This fact suggests the expansion is not effective for $\Upsilon (3S\to 1S)$. 
}  
not applicable to the case with the larger energy release. 

In the following we examine the consistency of our results,
especially the threshold behavior of $\Upsilon (3S\to 1S)$, 
with constraints from chiral symmetry,
and it is shown to be actually satisfied.

\section{Effective Lagrangian}
\subsection{Suppression behavior in linear $\sigma$ model} 
First we study the threshold suppression of $\pi\pi$ scattering amplitude by 
SU(2) linear $\sigma$ model (L$\sigma$M).
Through the mechanism of spontaneous chiral symmetry breaking the 
$\sigma$ acquires 
a non-zero vacuum expectation value (VEV) $\sigma_0 =f_\pi$,
and the $\sigma\pi\pi$ coupling 
(${\cal L}_{\rm int}=-g_{\sigma\pi\pi}\sigma '{\mib \pi}^2$) appears.
The $g_{\sigma\pi\pi}$ and $\lambda$ 
(defined by ${\cal L}_{\rm int}=-\lambda({\mib \phi}^2)^2/4$) 
are related with $m_\sigma$ as
\begin{eqnarray}
\sigma = \sigma_0+\sigma ' ;\ \sigma_0=f_\pi .\ \ 
g_{\sigma\pi\pi}=f_\pi \lambda =(m_\sigma^2-m_\pi^2)/(2f_\pi ).
\label{eq3}
\end{eqnarray}
The $\pi\pi$ scattering $A(s,t,u)$ amplitude is given as the sum 
of terms, attractive $\sigma$ amplitude and repulsive $\lambda\phi^4$ amplitude.
According to Eq.(\ref{eq3}),
these two amplitudes strongly cancel with each other in $O(p^0)$ level(, where $p$ means
a momentum of the pion), leaving  the $O(p^2)$ Tomozawa-Weinberg amplitude, which is consistent with 
the derivative coupling property of Nambu-Goldstone $\pi$ meson.
\begin{eqnarray}
A(s,t,u) &=& \frac{(-2g_{\sigma\pi\pi})^2}{m_\sigma^2+(p_1+p_2)^2}-2\lambda
   =\frac{1}{f_\pi^2}\left[ 
\frac{(/ \hspace{-0.3cm}m_\sigma^2-m_\pi^2)^2}
{(/ \hspace{-0.3cm}m_\sigma^2+(p_1+p_2)^2} -(/ \hspace{-0.3cm}m_\sigma^2-m_\pi^2)
      \right]  \nonumber\\
 &=& \frac{(m_\sigma^2-m_\pi^2)(-(p_1+p_2)^2-m_\pi^2)}
{m_\sigma^2+(p_1+p_2)^2}
\approx \frac{-(p_1+p_2)^2-m_\pi^2}{f_\pi^2}, 
\label{eq22}
\end{eqnarray}
where $p_1,p_2$ are momenta of the emitted pions.
The final form has Adler 0: 
The $A(s,t,u)$ vanishes when $p_{1\mu}$ is continued as 
$p_{1\mu}\longrightarrow 0_\mu$ but $p_{2\mu}$ remains on mass shell.  
This corresponds to zero at $s=-(p_1+p_2)^2=m_\pi^2$, 
being close to the threshold.

\subsection{Effective $\Upsilon$ decay 
interaction--Non-derivative type}
The similar cancellation is obtained in the $\Upsilon$ decay amplitude 
derived by the effective chiral symmteric Lagrangian of non-derivative type,
\begin{eqnarray}
{\cal L}_{\rm prod} & = & 
\xi_{2\pi} \Upsilon '_\mu \Upsilon_\mu (\sigma^2+{\mib \pi}^2),
\label{eq23}
\end{eqnarray}
where $\Upsilon ' (\Upsilon )$ is the field of initial (final) 
$b\bar b$ quarkonium.
$\xi_{2\pi}$ is the direct $2\pi$ (and $2\sigma$) production coupling constant. 
Through the chiral symmetry breaking,
the direct one-$\sigma$ production coupling 
$({\cal L}_\sigma =\xi_\sigma \Upsilon '_\mu \Upsilon_\mu \sigma )$
appears, and the $\xi_\sigma$ is related with $\xi_{2\pi}$.  
\begin{eqnarray}
{\cal L}_{\rm prod} & = & 
\xi_{2\pi} \Upsilon '_\mu \Upsilon_\mu 
(f_\pi^2+2f_\pi\sigma'+\sigma'^2+{\mib \pi}^2),\ \ \ 
\xi_\sigma =2 f_\pi \xi_{2\pi}.
\label{eqxi}
\end{eqnarray} 

The $\pi\pi$ production amplitude ${\cal F}$ is given as
the sum of ${\cal F}_\sigma$ and ${\cal F}_{2\pi}$, which cancel with each other 
in $O(p^0)$ level due to
the constraint Eq.~(\ref{eqxi})  of $\xi_\sigma$ and $\xi_{2\pi}$,
\begin{eqnarray}
{\cal F} &=& {\cal F}_\sigma  + {\cal F}_{2\pi} = 
\frac{\xi_\sigma (-2g_{\sigma\pi\pi})}{m_\sigma^2-s} + 2\xi_{2\pi}
=2\xi_{2\pi}\left( 
- \frac{/ \hspace{-0.3cm}m_\sigma^2-m_\pi^2}{/ \hspace{-0.3cm}m_\sigma^2
-s}
+/ \hspace{-0.2cm}1 \right) 
= 2\xi_{2\pi}\frac{m_\pi^2 -s}{m_\sigma^2-s},
\nonumber
\end{eqnarray}  
where $s=-(p_1+p_2)^2$.
The final amplitude takes $O(p^2)$ form, being consistent with the derivative coupling property
of $\pi$ meson. 
The Adler 0 occurrs at $s=-(p_1+p_2)^2=m_\pi^2$.
This is consistent with the experimental threshold behavior
in $\Upsilon (2S) \to \Upsilon (1S) \pi\pi$,
while is not with that in 
$\Upsilon (3S) \to \Upsilon (1S) \pi\pi $.

\subsection{Effective $\Upsilon$ decay 
interaction--Derivative type}
In order to explain the threshold behavior of $\Upsilon (3S\to 1S)$ decay,
we consider the following chiral symmteric 
``derivative type" interaction.\footnote{
This interaction as well as Eq.(\ref{eq23})
is easily shown to lead to the $S$-wave dominance
of the $\pi\pi$ system, experimentally confirmed.\cite{rfSwave}
}
\begin{eqnarray}
{\cal L}_{\rm prod}^{(d)} & = & 
\xi_{2\pi}^{(d)} \partial_\lambda \Upsilon "_\mu \partial_\nu \Upsilon_\mu 
(\partial_\lambda\sigma\partial_\nu\sigma
+\partial_\lambda{\mib \pi}\cdot\partial_\nu {\mib \pi}).
\end{eqnarray}
Since this interaction is of derivative form, 
the mechanism of 
chiral symmetry breaking gives no one-$\sigma$ production coupling.
Thus, there is no ${\cal F}_\sigma$ amplitude cancelling ${\cal F}_{2\pi}$.
Then the $\pi\pi$ production amplitude is given by 
\begin{eqnarray}
{\cal F}^{(d)} &=& -\xi^{(d)}(P"\cdot p_1 P\cdot p_2+P"\cdot p_2 P\cdot p_1),
\label{eq11}
\end{eqnarray}
where $P"(P)$ is the momentum of $\Upsilon (3S)$($\Upsilon (1S)$). 

In the relevant $\Upsilon (3S\to 1S)$ decay, the relation,
\begin{eqnarray}
M_{\Upsilon (3S)} > M_{\Upsilon (1S)} \gg 
M_{\Upsilon (3S)} - M_{\Upsilon (1S)} (\equiv \Delta E) \gg m_\pi , 
\end{eqnarray}
holds, where $\Delta E (= 895$MeV) is the energy release of the relevant decay.
Thus, in the rest frame of the initial $\Upsilon (3S)$
the final $\Upsilon (1S)$ is almost at rest, while the emitted two pion system
accepts a large relativistic recoil velocity. 

The ${\cal F}^{(d)}$ vanishes
when $p_{1\mu}\to 0_\mu$. Thus this ${\cal F}^{(d)}$ 
has Adler 0, and satisfies the general constraints from chiral symmtery.
However, the limit, $p_{1\mu}\to 0_\mu$, is far from the momentum in the physical $\pi\pi$
threshold, and the corresponding 0 
does not lead to the suppression near the threshold,
since at $s=4m_\pi^2$ the pion four-momenta should be $p_{1\mu}=p_{2\mu}$ and 
the pion energy becomes 
$p_{10}=p_{20}\approx (M"-M)/2=450$MeV$\gg 0$.
Actually the ${\cal F}^{(d)}$ can be approximated
as ${\cal F}^{(d)}\approx -2\xi^{(d)}M"Mp_{10}p_{20}$,
which is almost $s$-independent in all the physical region 
and has no zero close to the threshold.
By using this type of amplitude we can explain the steep increase of the 
$\Upsilon (3S)\to\Upsilon (1S)\pi\pi$ spectra. 

In the analysis of Part I we have used the
free parameters 
$r_\sigma$ and $r_{2\pi}$.
They correspond to the $\xi_\sigma$ and $\xi_{2\pi}$ in the effective Lagrangian,
which 
have no constraints from chiral symmetry,  
and our treatment is proved to be correct.

\section{Comparison between situations of $\pi\pi$ production 
and scattering}
Summarizing our considerations given in the previous sections, we
compare the general features of $\pi\pi$ production processes with those 
of $\pi\pi$ scattering process in Table I.

\begin{table}
\caption{Comparison of $\pi\pi$ production with $\pi\pi$ scattering}
\begin{tabular}{|l|l|l|}
\hline
                 & $\pi\pi$ production & $\pi\pi$ scattering\\
\hline
Energy release $\Delta E$ & $\gg m_\pi$, generally large & $\approx 0$, close to threshold \\
\underline{Chiral momentum expansion} & \underline{Not valid} & \underline{valid}\\
\hline
Form of amplitude & $P\cdot p_1 P\cdot p_2$ & $p_3\cdot p_1$ etc. \\
Amplitude near threshold &  $\sim O(M \Delta E)^2$ large & $\sim O(m_\pi^2)$ small \\
\hline 
Cancellation of ${\cal F}_\sigma$ and ${\cal F}_{2\pi}$ & generally No & Yes \\
\ \ \ near $\pi\pi$ threshold & & \\
Adler 0 limit $p_{1\mu}\to 0_\mu$ & far from thres. region 
                         & close to thres. region\\
\hline
Feature of spectra & Steep increase from threshold & Suppresion near threshold  \\
     & Direct $\sigma$ peak in many cases & No direct $\sigma$ peak \\
\hline
\end{tabular}
\end{table}

The $\pi\pi$ production processes have generally much the larger $\Delta E$ 
than $m_\pi$. Thus, the momentum of emitted pion becomes large, and the
chiral momentum expansion in  
nonlinear treatment of pion is not applicable.
The derivative type amplitude, as Eq.(\ref{eq11}), 
may play an important role. The Adler 0 exists
but its limit $p_{1\mu}\to 0_\mu$ is far from the physical momentum 
in threshold region, and so 
it does not give the suppression at 
the small $s$ region, the spectra
shows steep increase from threshold, and the direct $\sigma$ peak
is observed in $m_{\pi\pi}\approx 500\sim 600$MeV.
 
On the other hand in the $\pi\pi$ scattering process, the $\Delta E\approx 0$ 
corresponds to the near-threshold region, and pion momentum itself becomes small.
So the chiral perturbation theory is effective in this case,
and 
the spectra are suppressed close to threshold.

We can see the above mentioned situations actually in the following examples:\cite{rf8} 
In the $pp$-central collision experiment, $pp\to pp (\pi^0\pi^0)$, by GAMS
the large event concentration in the low energy region $m_{\pi\pi}\approx 500$MeV is
explained as due to direct production of $\sigma$ resonance.
The initial fast proton has large momentum of 450 GeV/c, and this process has
very large $\Delta E$ giving the pion extremely high momentum
in $\pi\pi$ threshold. 
In $J/\psi \to \omega\pi\pi$ decay,
where $\Delta E$ is very large, 
the chiral cancellation does not occur,
and the direct $\sigma$ peak was observed directly. 

In the relevant problem 
the $\Delta E$ are  
$( \Delta E_{\Upsilon (3S\to 1S)},
\Delta E_{\Upsilon (2S\to 1S)},
\Delta E_{\Upsilon (3S\to 2S)};$
$\Delta E_{\psi (2S\to 1S)})=(895,563,332,589)$MeV.
Among these values the $\Delta E_{\Upsilon (3S\to 1S)}$ is the largest.
Actually only this process shows the steep increase from threshold.

\section{Concluding Remarks} 
Through the above theoretical considerations we may conclude that  
the method of our analysis applied in Part I,
is shown to be consistent with the general requirement from chiral symmetry.
In the $\pi\pi$ production processes 
with large energy release to the $\pi\pi$ system,
the chiral cancellation between $\sigma$ amplitude and $2\pi$ amplitude 
near the $\pi\pi$ threshold does not occur,
and the direct $\sigma$ peak is generally observed, while 
in the $\pi\pi$ scattering with small energy release near the threshold 
the cancellation occurrs, and the direct $\sigma$ peak cannot be observed 
in the spectra.      

Finally we give a short comment on the possible origin of ${\cal L}^{(d)}$.
If we have the coupling to intermediate tensor glueball states 
$G_{\mu\nu}$, 
${\cal L}^G = \xi_g \partial_\lambda \Upsilon "_\mu \partial_\nu \Upsilon_\mu G_{\lambda\nu}$,
the ${\cal L}^{(d)}$ is obtained. 
This form of ${\cal L}^G$ is naturally derived from the
calculation in a covariant level classification scheme.\cite{rf7}

The other possible origin is the $B\bar B$ coupled channel 
effect,\footnote{
The coupled channel amplitude, originally considered 
by Lipkin and Tuan,\cite{rf10}
was used as the additional term to
the multipole expansion amplitude by Moxhay,\cite{rf11} 
to explain the spectra of $\Upsilon (3S)\to \Upsilon (1S)\pi\pi$. 
However, according to the theoretical estimate\cite{rf12} by using NRQM
the contribution due to the $^3P_0$ scalar state 
is not suffucient to explain the experimental data.
We consider the $B^*$ should be taken as a chiral particle.\cite{rf13}
}
$\Upsilon (3S)\to B\bar B\to \pi B^*\bar B \to \pi\pi B\bar B 
\to \pi\pi \Upsilon (1S)$,
which is expected to be strong 
since the threshold energy $2m_B$ is very close to
$m_{\Upsilon (3S)}$.
If $B^*$ is a scalar meson,  
the corresponding amplitude is 
proportional to the emitted pion energy,
and is similar to Eq.(\ref{eq11}).


\vspace*{-0.4cm}

\begin{thebibliography}{9}
\vspace*{-0.3cm}
\bibitem{rf1}T.~Komada, this proceedings. 
\bibitem{rf2}K. Gottfried, Phys. Rev. Lett. {\bf 40} (1978), 598. 
Y. P. Kuang and T. M. Yan, Phys. Rev. {\bf D24} (1981), 2874.
M. B. Voloshin, Nucl. Phys. {\bf B154} (1979), 365. 
\bibitem{rf3}M. B. Voloshin and V. Zakharov, Phys. Rev. Lett. {\bf 45} (1980), 688.
V. A. Novikov and M. A. Shifman, Z. Phys. {\bf C8} (1981), 43. 
\bibitem{rf5}P. Moxhay and J. P. Rosner, Phys. Rev. {\bf D28} (1983), 1132.
\bibitem{rf6}G. Belanger, T. DeGrand and P. Moxhay, Phys. Rev. {\bf D39} (1989), 257.
\bibitem{rfSwave}F. Butler et al. (CLEO), Phys. Rev. {\bf D49} (1994), 40.
\bibitem{rf7}S. Ishida, this proceedings; in proc. of Hadron97 in BNL, 1997, 
ed. by S.U. Chung and H.J. Willutzki, AIP conf. proc. 432; 
in proc. of WHS99 in Frascati, 1999,
ed. by T.Bressani, A. Feliciello and A. Filippi, Frascati Physics Series 15, 1999. 
\bibitem{rf8}K. Takamatsu, in proc. of Hadron97 at BNL, 1997.
\bibitem{rf10}H. J. Lipkin and S. F. Tuan, Phys. Lett. {\bf B206} (1988), 349. 
\bibitem{rf11}P. Moxhay, Phys. Rev. {\bf D39} (1989), 3497.
\bibitem{rf12}H. Y. Zhou and Y. P. Kuang, Phys. Rev. {\bf D44} (1991), 756.
\bibitem{rf13} S. Ishida, M. Ishida and T. Maeda, Prog. Theor. Phys. {\bf 104} (2000), No.4; 
this proceedings.
\end{thebibliography}
\end{document}